\documentclass[useAMS,usenatbib]{mn2e}
\usepackage{epsf}
\usepackage{natbib}
\usepackage{amssymb}
\usepackage{graphicx}
\usepackage[usenames]{color}
\usepackage{fmtcount}

\newcommand {\bc}{\begin {center}}
\newcommand {\ec}{\end {center}}
\newcommand {\be}{\begin {equation}}
\newcommand {\ee}{\end {equation}}
\newcommand {\beq}{\begin {eqnarray}}
\newcommand {\eeq}{\end {eqnarray}}

\def\disp {\displaystyle}

\title[Resonant scattering spectral models]{Resonant scattering in the Perseus Cluster:  spectral model for constraining gas motions with {\it Astro-H}}
\author[Zhuravleva et al.]{I. Zhuravleva$^{1,2,3}$\thanks{zhur@stanford.edu}, E. Churazov$^{3,4}$, R. Sunyaev$^{3,4}$, S. Sazonov$^{4,5}$, S. W. Allen$^{1,2,6}$, \newauthor N. Werner$^{1,2}$, A. Simionescu$^{1,2,7}$, S. Konami$^8$, T. Ohashi$^8$\\
$^{1}$Kavli Institute for Particle Astrophysics and Cosmology, Stanford University, 452 Lomita Mall, Stanford, CA 94305-4085, USA\\
$^{2}$Department of Physics, Stanford University, 382 Via Pueblo Mall, Stanford, CA 94305-4060, USA\\
$^{3}$MPI f\"ur Astrophysik, Karl-Schwarzschild str. 1, Garching, 85741, Germany\\
$^{4}$Space Research Institute, Profsoyuznaya str. 84/32, Moscow, 117997, Russia\\
$^{5}$Moscow Institute of Physics and Technology, Institutsky per. 9, 141700 Dolgoprudny, Russia\\
$^{6}$SLAC National Accelerator Laboratory, 2575 Sand Hill Road, Menlo Park, CA 94025, USA\\
$^{7}$Institute of Space and Astronautical Science (ISAS), JAXA, 3-1-1 Yoshinodai, Chuo-ku, Sagamihara, Kanagawa 252-5210, Japan\\
$^{8}$Department of Physics, Tokyo Metropolitan University, 1-1 Minami-Osawa, Hachioji, Tokyo 192-0397, Japan
}

\begin{document}

\date{Accepted .... Received ...}

\pagerange{\pageref{firstpage}--\pageref{lastpage}} \pubyear{2013}

\maketitle

\label{firstpage}

\begin{abstract}
X-ray spectra from cores of galaxy clusters can be strongly distorted
by resonant scattering of line photons, affecting metal abundance and
gas velocity measurements. We introduce simulated spectral models
that take into account the resonant scattering effect, radial
variations of thermodynamic properties of the hot gas, projection
effects and small-scale isotropic gas motions. The key feature
of the models is that all these effects are treated self-consistently
for the whole spectrum, rather than for individual lines. The model
spectra are publicly available and can be used for direct comparison
with observed projected spectra. Comparison with the existing {\it
  XMM-Newton} and {\it Chandra} data of the Perseus Cluster shows
that even though there is no strong evidence for the resonant
scattering in Perseus, the low energy resolution of the X-ray CCDs is not
sufficient to robustly distinguish spectral distortions due to
the resonant scattering, different metal abundance profiles and different
levels of gas turbulence. Future {\it Astro-H} data will
 resolve most of the problems we are facing with CCDs. With the help
of our models, the resonant scattering analysis can be done
self-consistently using the whole spectral information, constraining
the level of gas turbulence already with a 100 ks observation
with {\it Astro-H}.
\end{abstract}

\section{Introduction}
\label{sec:intr}

The scattering of photons of optically thick lines, so-called resonant
scattering (hereafter RS), is important in a variety of astrophysical
conditions \citep[see e.g.,][for a review]{Chu10}. In the context of hot plasma in galaxy clusters the RS effect was first considered by
\citet{Gil87}. They showed that even though clusters are transparent
in their thermal continua, the optical depth $\tau$ in some resonant X-ray lines
can be of the order of few. Scattering of photons in the optically thick lines
significantly distorts both the radial surface brightness profile and the shape
of the lines.

\begin{figure}
\begin{minipage}{0.47\textwidth}
\includegraphics[trim=0 155 0 150,width=1\textwidth]{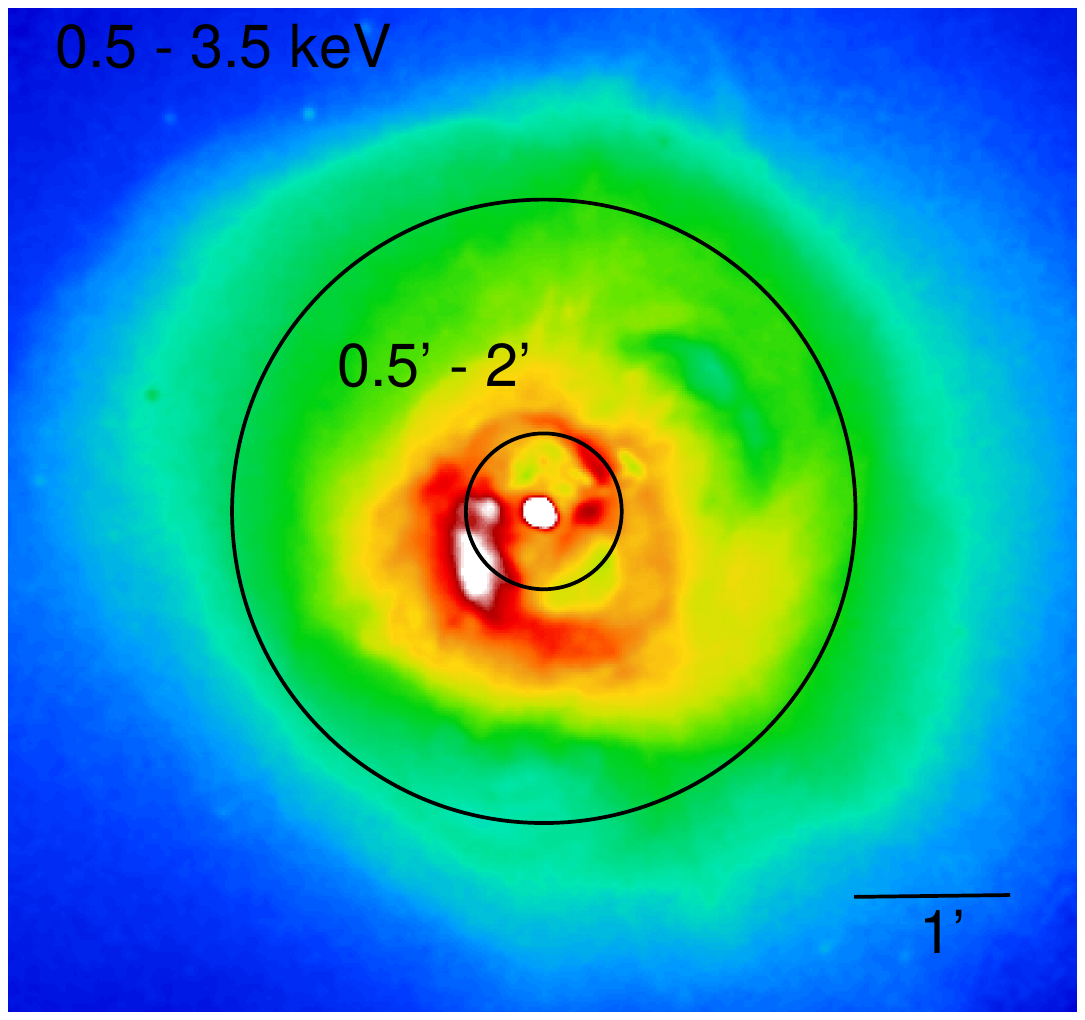}
\caption{
{\it Chandra} image of the central region of the Perseus Cluster in 0.5-3.5 keV energy band. The
image is slightly smoothed with a 3 arcsec Gaussian. The annulus 0.5-2 arcmin is indicated. This annulus is chosen to show results of resonant scattering effect in the cluster.
\label{fig:perim}
}
\end{minipage}
\end{figure}

Distortions in spectra and intensity due to the RS effect are less prominent if
turbulent motions of gas are present. Therefore, by measuring the level
of line flux suppression in the optically thick lines, one can constrain
the velocity amplitude of motions. This idea has been studied 
with radiative transfer simulations \citep[see
e.g.][]{Shi98,Chu04}. It has been shown that the scattering also produces
polarization of the same lines
\citep{Saz02} in the X-ray spectra of galaxy clusters, which provides a unique way
to probe gas motions perpendicular to the line of sight \citep{Zhu10}.

Detection of the RS is difficult with the current generation of X-ray
observatories due to the limited energy resolution of CCDs. For instance,
several attempts have been made to detect the RS signal in the Perseus Cluster using the 6.7 keV line of the He-like iron \citep{Mol98,Aki99,Chu04,Gas04,Eza01,Dup01,San04}. These studies revealed no strong evidence for the RS in the
data. However, the spectral lines are unresolved, and
uncertainties in the metal abundance profiles and projection effects
complicate firm conclusions on the level of turbulence.

 At lower energies ($\sim$ 0.5-2.5 keV), a better energy resolution can be reached using the Reflection Grating Spectrometer (RGS) measurements on {\it XMM-Newton}. The spatial extent of the cooling core of the Perseus Cluster prevents any meaningful observational constraints on 
the velocity broadening of spectral lines using RGS. However, the RGS measurements have been used to place constraints on turbulence in more compact cores of elliptical galaxies and cool-core clusters by measuring directly the width of individual lines \citep[mostly upper limits, see e.g.][]{San11,San13,Bul12} and by using the RS analysis in the Ne-like FeXVII at 15$\rm \mathring{A}$ and 17$\rm \mathring{A}$ \citep[see e.g.][]{Xu02,Kah03,Wer09, deP12}. These studies found gas velocities (or upper limits) at the level of few 100 km/s. Although these lines are detected in the core of the Perseus Cluster, they are weak, and the emission in the lines is associated with the extended and complex network of filaments, which would be impossible to model realistically for the RS studies.

\begin{figure*}
\begin{minipage}{0.49\textwidth}
\includegraphics[trim=20 150 0 100,width=1\textwidth]{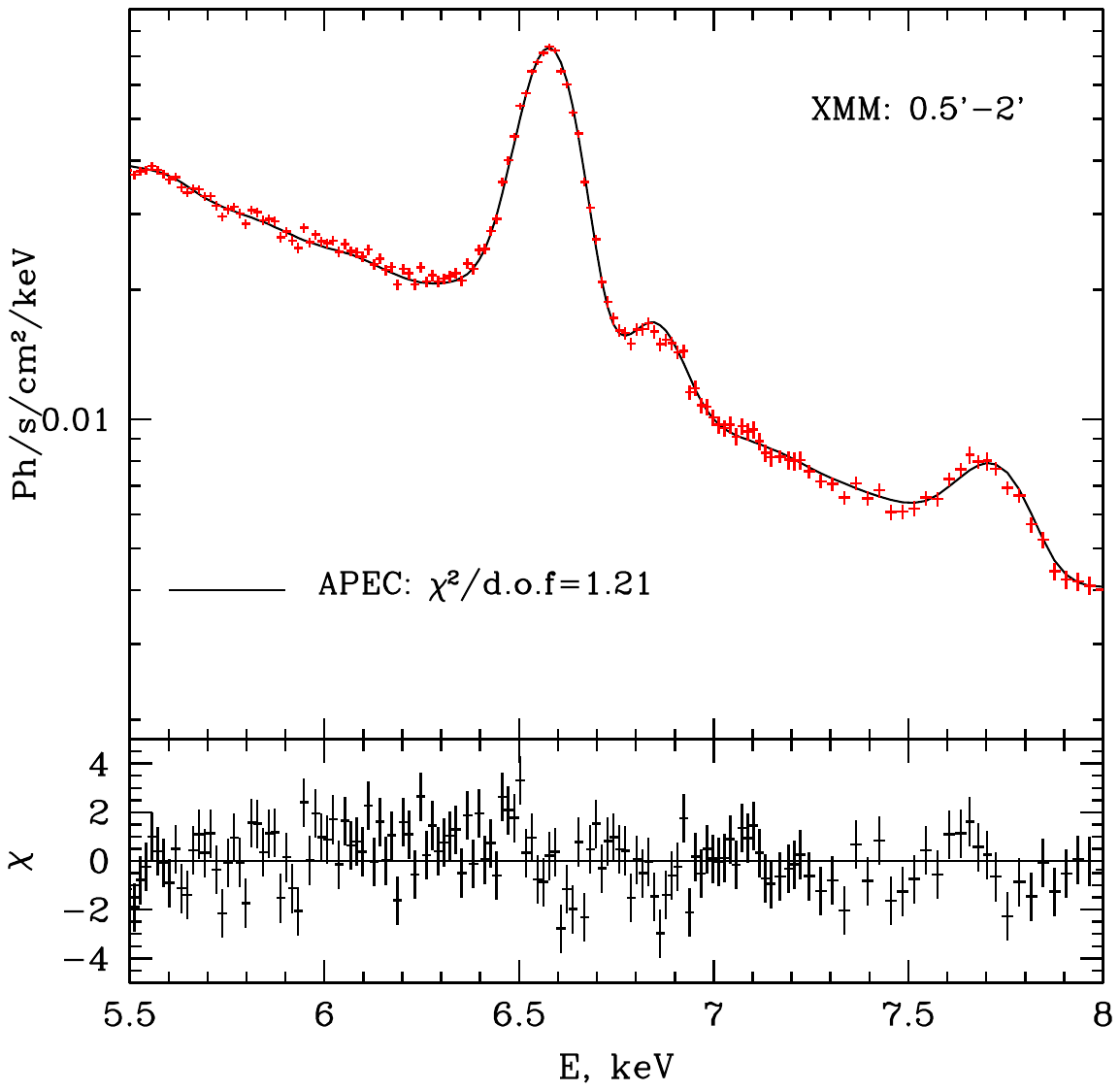}
\end{minipage}
\begin{minipage}{0.49\textwidth}
\includegraphics[trim=0 150 20 100,width=1\textwidth]{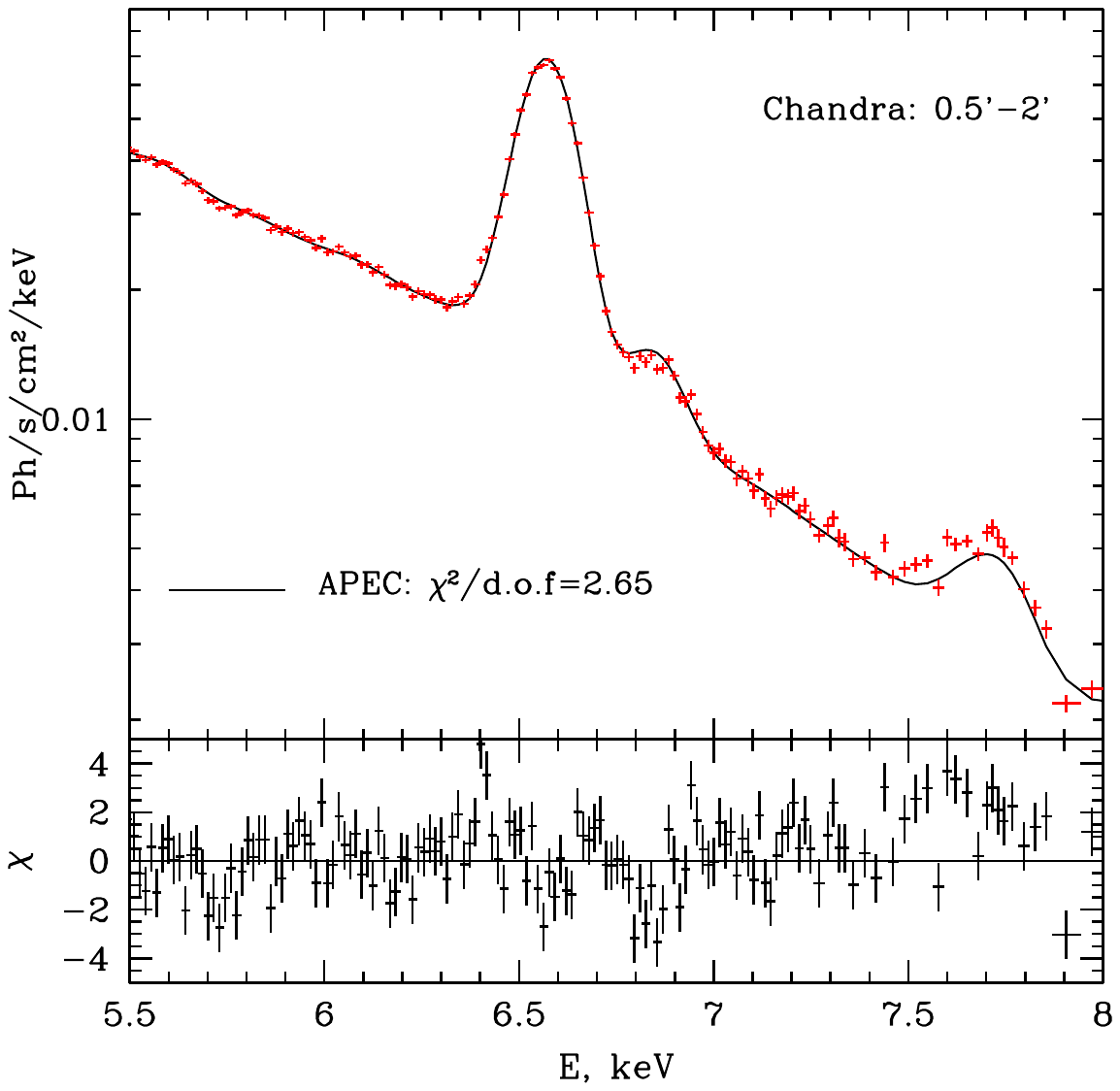}
\end{minipage}
\caption{Projected spectra (red points) of the Perseus Cluster from the central 0.5-2 arcmin annulus obtained from the {\it XMM-Newton} (left) and {\it Chandra} (right) observations. In both the cases data is fitted by APEC model in 5-9 keV energy band (black curve). Residuals are shown in lower panels.
\label{fig:rs_data_mod}
\label{fig:sp_proj}
}
\end{figure*}

To date the RS analysis has been based on comparisons of fluxes
in optically thick and thin lines. However, {\it Astro-H} will soon provide data with the energy
resolution $\sim$ 5 eV, allowing us for the first time to
resolve many individual lines in the spectra of galaxy clusters. Not only the line suppression can be detected, but also the modification of
the line shape induced by the RS. This will push the RS analyses to a new level of precision and make the constraints on the
turbulent velocities more robust.
  
The best way to study the RS is to fit the complete set of observed
data with spectral models that take into account projection
effects, line broadening due to thermal and turbulent motions of gas, and distortions in the line shape and flux due
to scattering. Here, we introduce simulations that produce such
models. Using emissivities in lines and continuum from detailed models of
the optically thin plasmas, we calculate the RS effect within given energy
bands using a Monte Carlo approach, producing projected model spectra in the
format of XSPEC models \citep{Arn96}. The models take into account radial
variations of density, temperature and abundance of heavy elements,
partially solving the problem of the presence of multi-temperature (in
projection) plasma components. The models are produced assuming different levels of
turbulence. Using these models, analysis of the RS effect within given energy band can be done simultaneously and self-consistently, taking into account all lines in the spectra. To illustrate the application of our spectral models, we use available observations of the Perseus Cluster. We also discuss future application of the models to high spectral resolution data from {\it Astro-H}.

The structure of the paper is as follows. The data analysis and 
 a 3D model of the Perseus Cluster are described in Section
2. Our radiative transfer simulations are presented in Section
3. Examples of our spectral models that account for the RS, projection effects and gas turbulence, are shown
in Section 4.1. Application of these models on existing data, and
to future data from {\it Astro-H} are discussed in Sections 4.2 and 4.3
respectively. Our conclusions are summarized in Section 5.

\section{X-ray data analysis and parameterization of the Perseus Cluster properties}
\label{sec:mod}

In order to perform simulations of spectra with RS, we adopt a spherically-symmetric model of a cluster based on observations of the Perseus Cluster with {\it XMM-Newton} and {\it Chandra} to obtain the radial dependence of the electron number
 density, temperature and abundance of heavy elements. In particular, we use all available archival data from the EPIC-MOS CCD array
 on {\it XMM-Newton} with a total exposure $\sim ~580$ ks, and the {\it Chandra} ACIS-S
observations with a total exposure $\sim 780$ ks. Calibrated event lists from the {\it XMM} observations
were generated using SAS version 12.0.1
\citep[the details of the analysis are described by][]{Chu03}. The {\it
Chandra} data are processed using the latest calibration data and
following the algorithms developed by
\citet{Vik05}. Fig. \ref{fig:perim} shows the {\it Chandra} image of the Perseus Cluster.

Excluding point sources and a 12 arcsec region around the central AGN, the projected spectra in radial annuli are obtained. Fig. \ref{fig:sp_proj} shows the projected spectra from the 0.5-2 arcmin annulus. We fit the projected spectra with XSPEC (version 12.8.0) using 1T APEC plasma models with AtomDB version 2.0.1 \citep{Smi01,Fos12} leaving temperature, metal abundance and redshift as free parameters. When fitting the spectra in a broad 0.5-9 keV band, the fit is poor, even if we use two-temperature (2T) plasma model. Reduced $\chi^2$ and the best-fitting temperatures and abundances are shown in Table \ref{tab:chi2}. This is not surprising, since the spectrum is multi-temperature both due to projection effects and due to the presence of the ICM with different temperatures at the same spatial distance from the cluster center. In order to avoid strong bias in the derived parameters, we restrict our analysis to the harder band 5-9 keV, where the role of cool gas is less significant. This band contains the 6.7 keV Fe line complex, where the RS is the strongest for Perseus. Fig. \ref{fig:sp_proj} shows the best-fitting 1T APEC models in the 5-9 keV band plotted on top of the observed data.  Note, that in this band the 1T model represents data quite well. Table \ref{tab:chi2} shows the goodness of the best-fitting models (1T-model, 2T-model and a model with variable Ni abundance). Notice, that changing abundance to \citet{Lod03}, allowing for 2T plasma or varying Ni abundance do not give significant advantage over the 1T plasma models. $\chi^2$ remains unchanged for the 5-9 keV band.

\begin{table*}
 \centering 
\caption{Goodness of the best-fitting models of optically thin plasma to the observed projected spectra in the 0.5-2 arcmin annulus in the Perseus Cluster. Spectra from the {\it XMM-Newton} and {\it Chandra} observatories are fitted in a broad 0.5-9 keV band and harder 5-9 keV band. $\chi^2$ is shown for one-temperature APEC models assuming two difference tables of Solar abundances \citep[A\&G and Lodd respectively,][]{And89,Lod03}, two-temperature APEC model and VAPEC with the Ni abundance as an additional free parameter.  In the last model, metal abundances of all elements are tied except for the Ni. The best-fitting temperatures and abundances of Fe and Ni are also listed.}
\begin{tabular}{@{}cccccccccccc@{}} 
\hline 
Model &\multicolumn{4}{c}{{\bf 0.5-9 keV}}&&\multicolumn{4}{c}{{\bf 5-9 keV}}\\
& T$_1$,&T$_2$,&Z$_{Fe}$/Z$_{\odot}$&Z$_{Ni}$/Z$_{\odot}$&$\chi ^2$/d.o.f.& &T$_1$,&T$_2$,&Z$_{Fe}$/Z$_{\odot}$&Z$_{Ni}$/Z$_{\odot}$&$\chi ^2$/d.o.f.\\
& keV & keV & & &(d.o.f.)& & keV & keV & & & (d.o.f.)\\
\hline
{\bf XMM-Newton}& &&&&&&&\\
A\&G, 1T & 3.59 & - & 0.50 & - & 23.05 (561)& & 4.43 & - & 0.48 & - & 1.20 (261)\\
Lodd, 1T & 3.96 & - & 0.9 & - & 35.14 (561) & & 4.37 & - & 0.72 & - & 1.19 (261)\\
2T & 2.17 & 5.01 & 0.45 & - & 13.77 (559)& & 4.3  & 64.0 & 0.48 & - & 1.18 (259)\\
vapec, Ni & 3.59 & - & 0.51 & 0.86 & 22.9 (560)& & 4.47 & - & 0.50 & 0.59 & 1.20 (260)\\
\hline
{\bf Chandra}& &&&&&&&\\
A\&G, 1T& 3.58& - & 0.52 & - & 75.42 (577) & & 4.33 & - &0.47 & - & 2.65 (269)\\
Lodd, 1T & 4.01 & - & 0.99 & - & 59.4 (577) & & 4.31 & - & 0.72 & - & 2.51 (269)\\
2T& 5.23 & 2.17 & 0.44 & - & 53.37 (575)& & 3.45 &22.5 & 0.58 & - & 2.60 (267)\\
vapec, Ni & 3.57 & - & 0.51 & 0 & 74.12 (576)& & 4.30 & - & 0.47 & 1.02 & 2.49 (268)\\
\hline
\label{tab:chi2}
\end{tabular}
\end{table*}

 The success of the  1T APEC model with the {\it XMM-Newton} projected spectrum in the core of
the cluster in the 5-9 keV band is particularly interesting. We do not see
suppression in the K$_\alpha$ FeXXVI line complex due to the RS. This suggests
that turbulent motions of gas are present in the core of Perseus. This
conclusion is in agreement with previous findings of
\citet{Mol98,Chu04, Tam09}.  
However, comparing the {\it XMM-Newton} and {\it Chandra} spectra we see
that the 1T APEC model visibly deviates from the {\it Chandra} spectrum
near the 7.8 keV feature. We attribute this discrepancy to the imperfect
cross-calibration of both instruments\footnote{For instance, see the
  discussion of cross-calibration issues at the 8th IACHEC meeting in
  UK: http://web.mit.edu/iachec/meetings/}.  Our experiments with various
versions of the APEC code (version 1.3.1 -- 2.0.1) show that 
differences of the order of few per cent are possible in the energy
range of interest.  The uncertainties in the
plasma emission models, e.g., in the ionization balance, might give
additional non-negligible contributions \citep{Fos12,Loe12}. While these small differences are typically
unimportant, in the context of the RS effect they may play a role, since
the expected line suppression in low resolution spectra is of the order of
10 per cent or smaller. At the same time, statistics of the spectra,
accumulated by {\it Chandra} and {\it XMM-Newton} is sufficient to
detect per cent level deviations from the model. These calibration
uncertainties do not affect strongly the best-fitting parameters of
the 1T APEC model. 


The success of the 1T model in the 5-9
keV band  motivated us to use this band\footnote{We
  have checked that 1T APEC model in the broader 0.5-9 keV band
  still provides poor fit even to the deprojected spectra, although better
  than for the projected spectra. This indicates that at some level
  multi-temperature plasma is also present at the same 3D distance
  from the center.} to build a 3D spherically-symmetric model for the
Perseus Cluster by deprojecting observed spectra using the procedure
described in \citet{Chu03}. This model will be used to produce
spectral models that account for the RS, turbulent motions of gas, radial
variations of thermodynamic quantities and projection effects.
\begin{figure}
\includegraphics[trim= 15 150 0 100,width=0.5\textwidth]{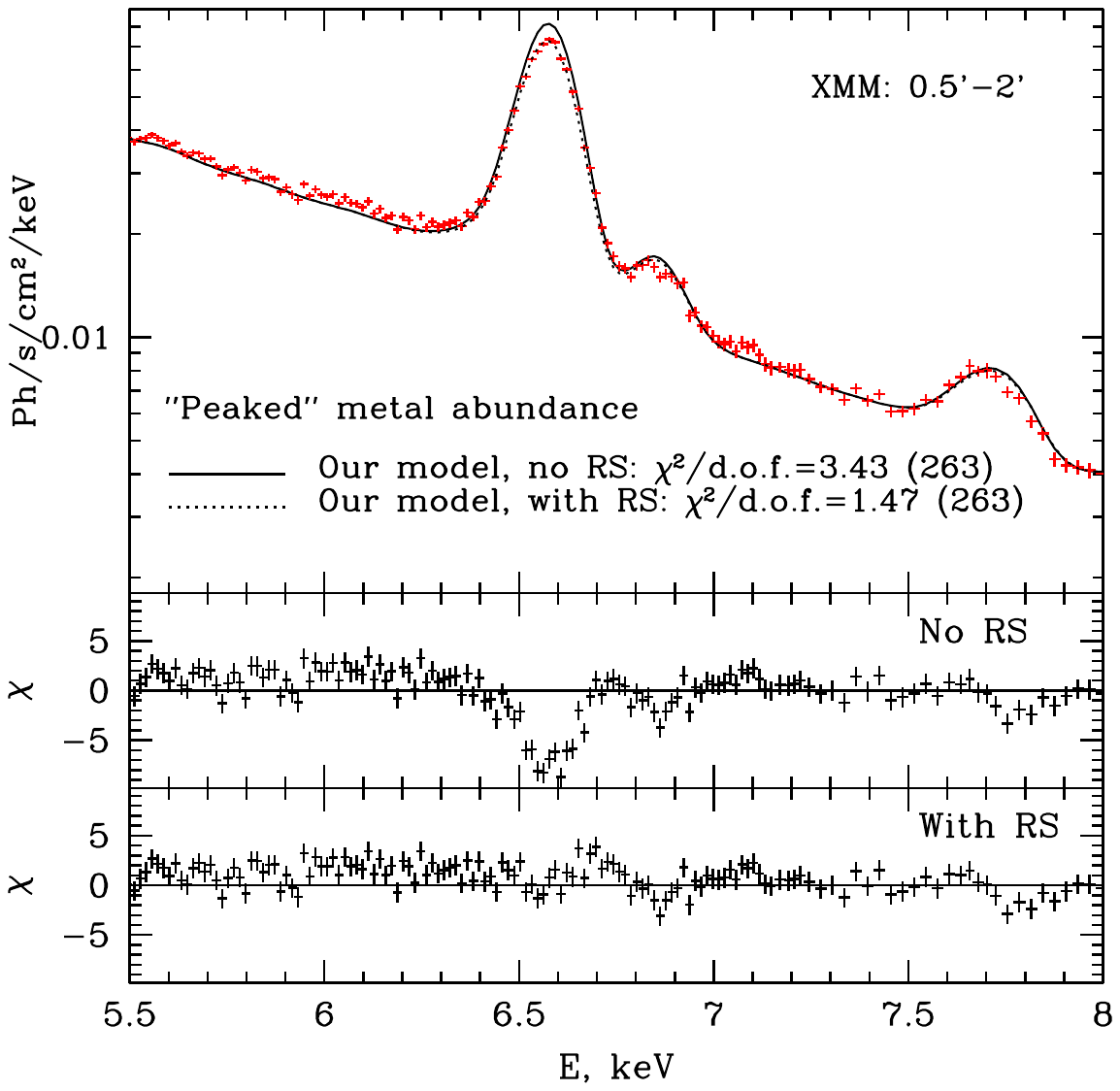}
\caption{ Deprojected number density of electrons (top), gas
  temperature (middle) and abundance of heavy elements (bottom) in the
  Perseus Cluster obtained from the {\it Chandra} and {\it XMM-Newton}
  observations in the $5-9$ keV energy band. Profiles were obtained
  while treating abundance of heavy elements both as a free parameter
  and fixing it to a constant value 0.5 relative to solar. Black solid
  curves show the approximation used as an input for our calculations
  of resonant scattering models. The dotted black curve shows the
  ``peaked-in-the-core'' metal abundance radial profile. Parametric
  fits are given in Appendix \ref{sec:appen}.  Vertical dotted lines
  across all panels indicate 0.5-2 arcmin region.
\label{fig:deproj}
}
\end{figure}

Deprojected profiles of electron number density, temperature and metal
abundance are shown in Fig. \ref{fig:deproj}. Results from both
observatories are consistent with each other although the scatter in
$T$ and $Z$ increases toward the cluster core. As starting values we
use the redshift $z=0.0176$ of the central galaxy NGC1275 from
NED\footnote{http://ned.ipac.caltech.edu/} and $N_H$ column density
$1.38\cdot 10^{21}$ cm$^{-2}$ in the core of the cluster
\citep{Kal05,Dic90}. $N_H$ is fixed, but the redshift is allowed to
vary. Given extremely high statistics, even very small variations of
the redshift, at the level of $\Delta z \sim 10^{-3}$ would cause
visible residuals in the spectra. The best-fitting values of the redshift
for the 1T APEC models are 0.0152 and 0.0167 for {\it XMM-Newton} and
{\it Chandra}, respectively. We assume that these variations are within
calibration uncertainties and use the best-fitting values. The
observed profiles are approximated with the smooth functions shown
with black curves in Fig. \ref{fig:deproj}.

Deprojected spectra were fitted with 1T APEC plasma model assuming
either constant metal abundance of 0.5 solar relative to
\citet{And89}, or leaving abundance as a free parameter of the
model. In terms of density and temperature profiles the assumption of
constant or variable abundance does not have strong impact. We
therefore can fix the density and temperature profiles and consider
several different models of the abundance radial profile. Contrary to
expectations of the highest abundance of heavy elements in the very
core of the cluster due to the presence of the massive elliptical
galaxy NGC1275, we see a drop of abundance in the core of Perseus,
broadly discussed in the literature \citep[see,
  e.g.,][]{Sch02,Chu03}. Similar drops are observed in other clusters
and are (at least in some systems) attributed to the use of
oversimplified spectral models in the presence of several temperature
components \citep[e.g.,][]{Buo98,Wer08}. RS can also be responsible
for the apparent drop of the abundance in the cluster core.  However,
neither of these explanations seems to be sufficient to fully
explain the metal abundance dip in the core of Perseus
\citep[see][]{San04, San06, Chu04,Tam09}. Given that the RS effect could
contribute to the drop of the abundance, we also perform simulations using
a "peaked'' abundance profile (black dotted curve in
Fig. \ref{fig:deproj}) in the core. This ``peaked'' profile was
selected to approximately compensate the suppression of the resonant
line flux by the RS (see Section \ref{sec:compar}). Functional
forms of $n_e$, $T$ and $Z$ are presented in Appendix A.

\begin{table*}
 \centering 
\caption{ The strongest lines
 marked in Fig. \ref{fig:rs_lines} and Fig. \ref{fig:rs_lines_6.7}: their
 energy, transition information (level notations are according to APEC), and the radial optical depth calculated at the center of line
 assuming two different metal abundance models.}
\begin{tabular}{@{}lcccrclccl@{}} \hline Ion & Energy, & Upper &
 Lower & \multicolumn{3}{c}{Transition}&$\tau (M=0)$ & $\tau (M=0)$&Comments \\ & keV & level & level & & & & ``free" ab.
 & ``peaked" ab.&\\
&&&&&&&eq. (A3) & eq. (A4)&\\
\hline
FeXXV  & 8.347 & 10192 & 3 & $ 1s2s(^1S_0)$&-&                             &  0.0e0 &  0.0e0       & satellite  \\
FeXXV  & 8.302 &  21      & 1 & $ 1s^2(^1S_0) $&   - &$1s4p(^3P_1)    $  &  9.3e-3 &  1.2e-2  &  He-like K$_{\gamma}$, intercomb.\\
FeXXV  & 8.293 & 23       & 1 & $ 1s^2(^1S_0)  $&  - &$1s4p(^1P_1)    $  &  1.1e-1 &  1.3e-1&  He-like K$_{\gamma}$, reson.\\
FeXXVI & 8.253 & 7         & 1 & $ 1s(^2S_{1/2}) $&- &$3p(^2P_{3/2})$  &  8.3e-2 &  9.7e-2& H-like K$_{\beta}$\\
FeXXVI & 8.246 & 6         & 1 & $ 1s(^2S_{1/2}) $&- &$3p(^2P_{1/2})$  &  4.1e-2  &  4.8e-2& H-like K$_{\beta}$\\
FeXXV  & 8.168 &  10127 & 3 & $ 1s2s(^1S_0)  $& - &                         &  0.0e0 &  0.0e0         & satellite   \\
FeXXV  & 7.882 & 11       & 1 & $1s^2(^1S_0)  $&  - &$1s3p(^3P_1)    $   &  2.8e-2 &  3.4e-2 & He-like K$_{\beta}$ , intercomb.\\ 
NiXXVII&7.806 &  7         & 1 & $1s^2(^1S_0)  $&  - &$1s2p(^1P_1)$      &  7.1e-2 &  8.8e-2& He-like K$_{\alpha}$, reson.   \\
NiXXVII& 7.799&  6         & 1 & $1s^2(^1S_0)  $&  - &$1s2p(^3P_2)$      &  2.3e-6 &  2.8e-6 & He-like K$_{\alpha}$, intercomb. \\    
NiXXVII& 7.779& 10001   & 1 & $1s^2(^1S_0)  $& -  &                          & 0.0e0 &  0.0e0    & satellite    \\
NiXXVII& 7.766 & 5          & 1 & $1s^2(^1S_0)  $&  - &$1s2p(^3P_1)$      & 7.7e-3 &  9.4e-3  & He-like K$_{\alpha}$ , intercomb.\\   
NiXXVII &7.744 &  2         & 1 & $1s^2(^1S_0)  $&  - &$1s2s(^3S_1)$       & 4.9e-8 &  6.0e-8 & He-like K$_{\alpha}$, forbid. \\
FeXXVI  & 6.973 & 4         & 1 & $1s(^2S_{1/2}) $& - &$2p(^2P_{3/2})$   &  2.6e-1 &  3.0e-1 & H-like K$_{\alpha}$ \\
FeXXVI  & 6.952 & 3         & 1 & $1s(^2S_{1/2}) $& - &$2p(^2P_{1/2})$   &  1.3e-1  &  1.5e-1  & H-like K$_{\alpha}$\\
FeXXVI  & 6.937 & 10017 & 6 & $3p(^2P_{1/2}) $& -  &                        &   0.0e0 &  0.0e0  & satellite  \\
FeXXVI  & 6.918 & 10004 & 2 & $2s(^2S_{1/2})  $& -  &                        &   0.0e0 &  0.0e0       & satellite  \\
FeXXV    & 6.700 & 7         & 1 & $1s^2(^1S_0)  $&  - &$1s2p(^1P_1)$      &   1.98e0 &  2.5e0 & He-like K$_{\alpha}$, reson.\\
FeXXV    & 6.682 & 6         & 1 & $1s^2(^1S_0)  $&  - &$1s2p(^3P_2)$      &	 4.6e-5 &  5.7e-5 & He-like K$_{\alpha}$, intercomb.\\ 
FeXXV    & 6.668 & 5         & 1 & $1s^2(^1S_0)  $&  - &$1s2p(^3P_1)$      &	 1.6e-1 & 2.0e-1 & He-like K$_{\alpha}$, intercomb. \\
FeXXV    & 6.655 &  10076 & 6 &$1s2p(^3P_2) $&  -   &                         &   0.0e0 & 0.0e0    & satellite  \\
FeXXV    & 6.637 &  2         & 1 &$1s^2(^1S_0)  $&  - &$1s2s(^3S_1)$      &	 8.4e-7 & 1.0e-6  &He-like K$_{\alpha}$, forbid.\\  
\hline
\label{tab:strong_lines}
\end{tabular}
\end{table*}

\begin{figure*}
\begin{minipage}{0.95\textwidth}
\includegraphics[trim=0 460 0 95,width=1\textwidth]{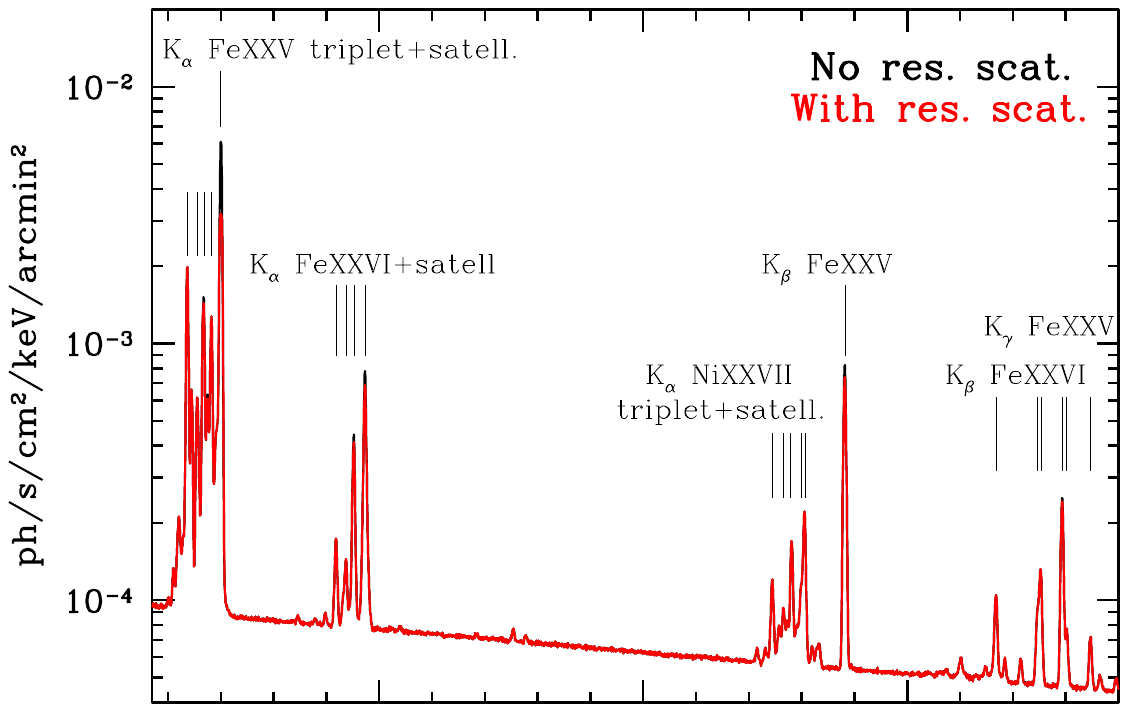}
\end{minipage}
\begin{minipage}{0.95\textwidth}
\includegraphics[trim=0 670 0 0,width=1\textwidth]{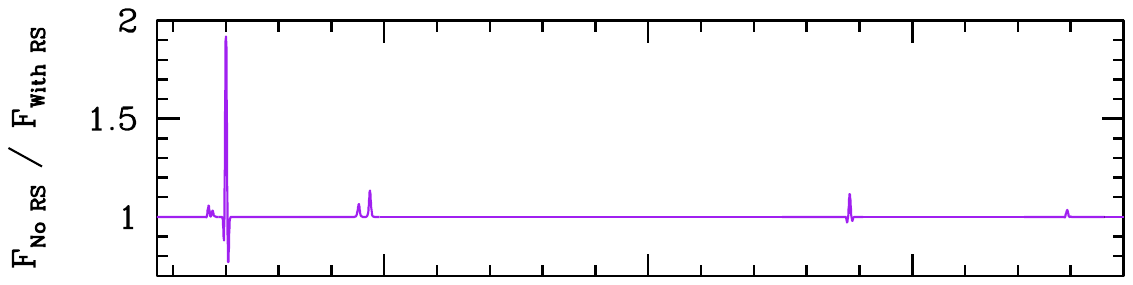}
\end{minipage}
\begin{minipage}{0.95\textwidth}
\includegraphics[trim=0 390 0 1,width=1\textwidth]{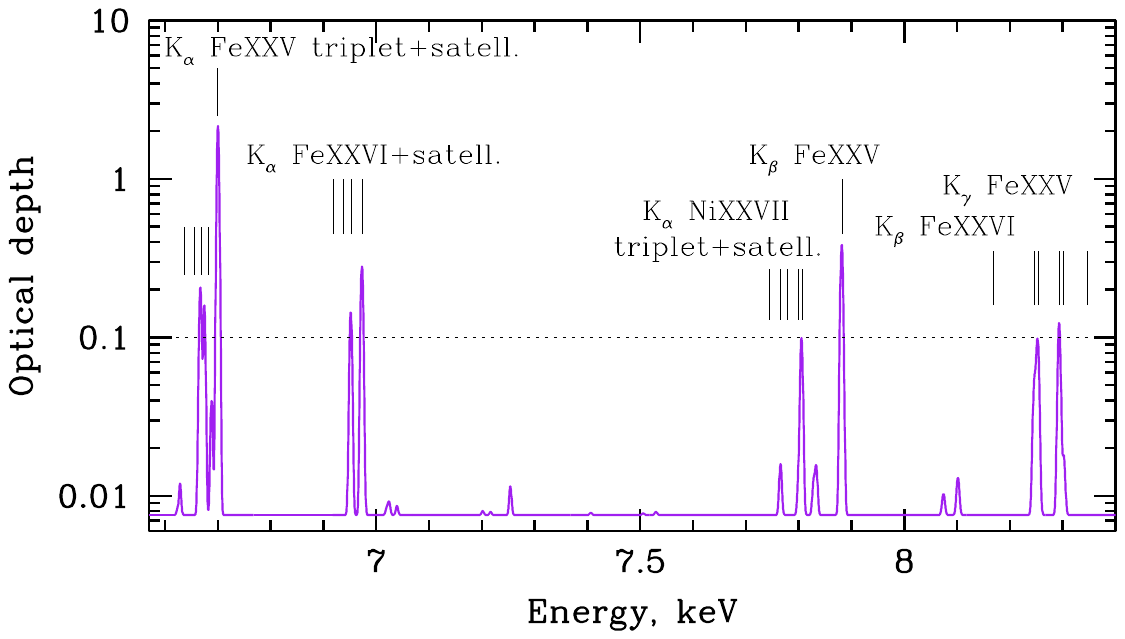}
\end{minipage}
\caption{
Predicted spectrum of the Perseus Cluster, which takes into account distortions of the line's flux and shape induced by resonant scattering.
{\bf Top:} projected spectra from the 0.5 - 2 arcmin annulus (see Fig. \ref{fig:perim}) both with (red) and without (black) the scattering. {\bf Middle:} ratio of spectral models without and with resonant scattering. {\bf Bottom:}
optical depth as a function of energy. Only lines with the optical depths above
the threshold $\tau=0.1$ (dotted line) are involved in the scattering process. The strongest lines are labeled and their characteristics are shown in Table \ref{tab:strong_lines}.
\label{fig:rs_lines}
}
\end{figure*}

\begin{figure*}
\begin{minipage}{0.49\textwidth}
\includegraphics[trim=10 150 30 90,width=1\textwidth]{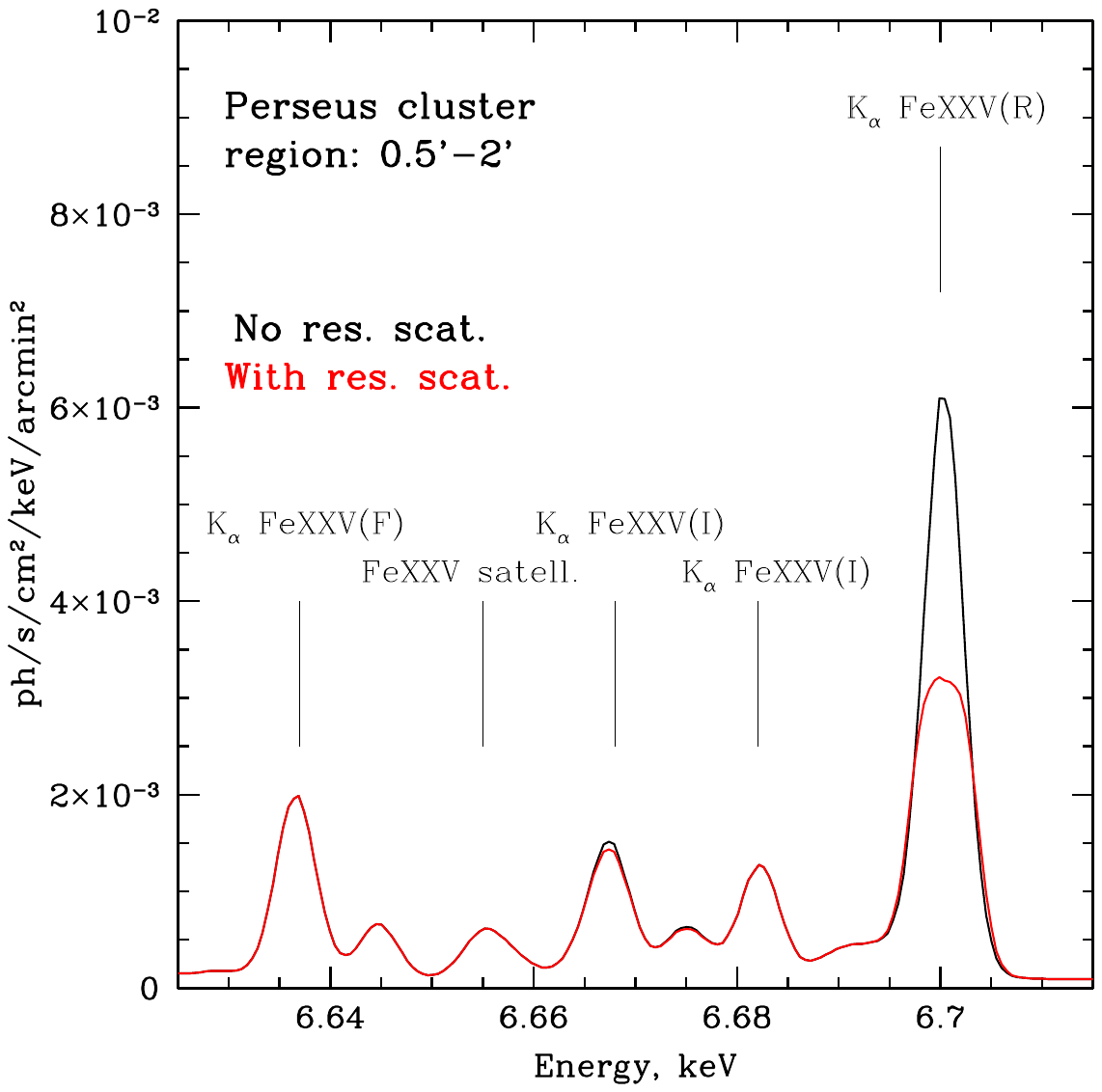}
\end{minipage}
\begin{minipage}{0.49\textwidth}
\includegraphics[trim=0 150 30 90,width=1\textwidth]{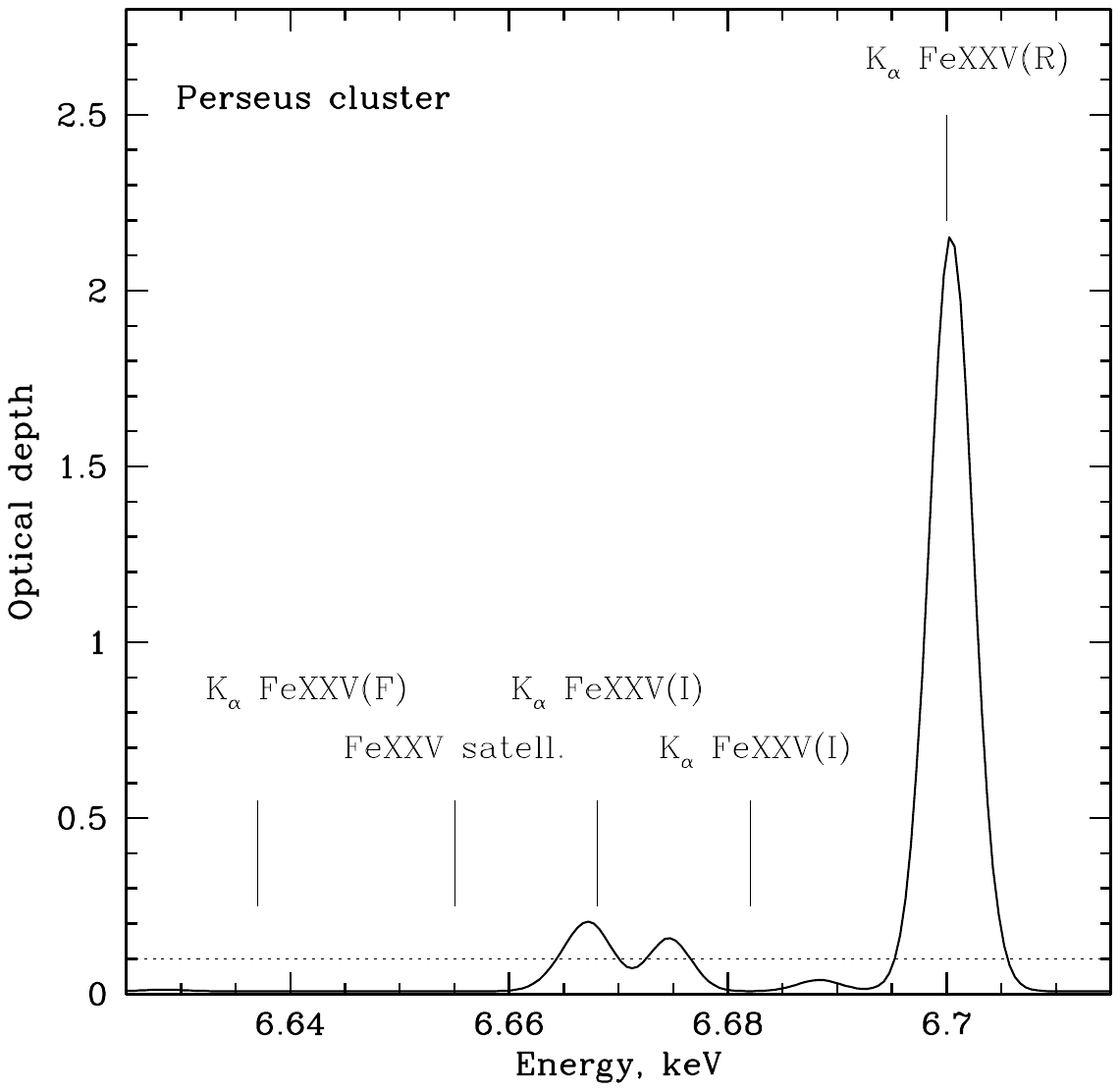}
\end{minipage}
\caption{
Resonant scattering effect in the He-like Fe triplet in the Perseus Cluster. See Fig. \ref{fig:rs_lines} and Table \ref{tab:strong_lines}.
\label{fig:rs_lines_6.7}
}
\end{figure*}

\section{Radiative transfer simulations}
\label{sec:sim}
The radiative transfer code developed for this work produces
projected spectra accounting for (i) the RS
effect, (ii) radial variations of density/temperature/abundance in
clusters and (iii) line broadening due to
the turbulent motions of gas. The code is based on earlier simulations
discussed in
\citet[][]{Saz02} and \citet{Zhu10,Zhu11}. The key feature of our code is the self-consistent treatment of the whole spectrum in the radiative transfer calculations. The output
 is in the same format of XSPEC models and allows one to directly
 compare the models with observed data in a usual
way. Below we briefly describe the code.

Once we have a cluster model (see Section \ref{sec:mod}), the spectrum
emitted by each spherical shell is calculated. The line emissivity, continuum and pseudo continuum are taken from APEC (AtomDB version 2.0.1) code \citep{Smi01,Fos12}. The continuum in APEC includes bremsstrahlung, radiative recombination and two photon emission, while pseudo continuum takes into account all weak lines which are not included in the emissivity files. Both the lines and
continuum associated with a given element are convolved with 
a Gaussian, to account for the line broadening. The Gaussian velocity width
 is determined both by thermal
broadening and broadening due to small-scale isotropic gas motions, as parameterized
through the Mach number\footnote{The Mach number is determined as a ratio of the 
 characteristic velocity of turbulence and the adiabatic
sound speed, $M=V_{\rm turb}/c_s$. Note, that in the code $V_{\rm
turb}\equiv \sqrt{2}V_{\rm 1D,turb}$, where $V_{1D,turb}$ is the velocity
dispersion along the line of sight. Coefficient $\sqrt{2}$ is chosen to be consistent with definitions from our previous works.} $M$
\be
\sigma=\frac{E_0}{\sqrt{2}}\left[\disp\frac{2kT_e}{Am_pc^2}(1+1.4AM^2)\right]^{1/2}.
\ee 
Here $E_0$ is the rest energy of the line, $A$ is the atomic mass of
the corresponding element, $m_p$ is the proton mass. Line broadening
due to radiative decay of levels is neglected. We use the definitions of solar abundances of
\citet{And89}. The ionization balance (collisional equilibrium) is
taken from \citet{Maz98}.

The optical depth from the center to the edge of cluster at the center of a line is defined as
\be
\tau=\int\limits_0^{\infty}\frac{\sqrt{2\pi} hr_ecf}{\sigma}n_pZ\delta_i dr,
\ee
where  $h$ is the Planck constant, $r_e$ is the classical radius of an electron, $f$ is the
oscillator strength of a transition, $n_p$ is the number density of protons, $Z$ is
the abundance of a given element relative to hydrogen and $\delta_i$
is the fraction of the elements in the appropriate ionization
state. The optical depth in continuum is calculated through the
Thomson cross section as follows
\be
\tau=\int\limits_0^{\infty}\sigma_T n_e dr.
\ee 
The scattering loop starts by drawing not only initial position and
direction of photon propagation, but also its energy. The photon has
initially a unit weight $w$ that reduces after each scattering by a 
factor $(1-e^{-\tau})$, where $\tau$ is the optical depth in the
direction of photon propagation. At the same time the quantity $w\cdot
e^{-\tau}$ is added to an appropriate bin (set by the energy and the
projected distance of the photon) of the emergent spectra. The scattering process repeats until
the weight drops below the set minimal value $w_{min}$, which in our
case is $w_{min}=10^{-6}$. During each scattering, the new position of the
emerging photon and the new direction are drawn in accordance with
the isotropic scattering phase matrix (the use of Rayleigh phase
function does not affect the results significantly unless polarization is taken into account). The photon
energy after the scattering is $E=E_0\left(1+\disp\frac{({\bf
V_{ion}m'})}{c}\right)$, where ${\bf V_{ion}}$ is the velocity of
the scattering ion, ${\bf m'}$ is the new direction of photon propagation
and $E_0$ is the line energy. The ion velocity in the direction of photon
propagation is $V_{ion,||}=c(1-E_0/E)$, so that the photon energy in
the reference frame of the ion is equal to $E_0$ and scattering
occurs. Other ion velocity components have Gaussian distribution with
the width set by the Doppler line width.

Since we consider scattering of photons in a certain energy band, a
large number of photons $n_{ph}$ is required in order to produce
statistically robust results (e.g., we use $n_{ph}=10^9$ for
simulations in 4-10 keV band). Such calculations are time consuming,
although straightforward parallelisation of the Monte Carlo
simulations can address this problem. Additionally, we introduce a
threshold in the optical depth $\tau_{min}$ of individual
lines. Only lines with $\tau > \tau_{min}=0.1$ participate in
the scattering process. To ensure good statistics over the whole
range of radii, a radially dependent weight is used to draw positions
of emitted photons. This allows one to reach robust results using
reasonable total number of photons. We refer the reader to our previous
works
\citep{Saz02,Zhu10,Zhu11} for more details on how the scattering process
is modeled in the simulations.

\begin{figure}
\begin{minipage}{0.48\textwidth}
\includegraphics[trim=20 150 20 100,width=1\textwidth]{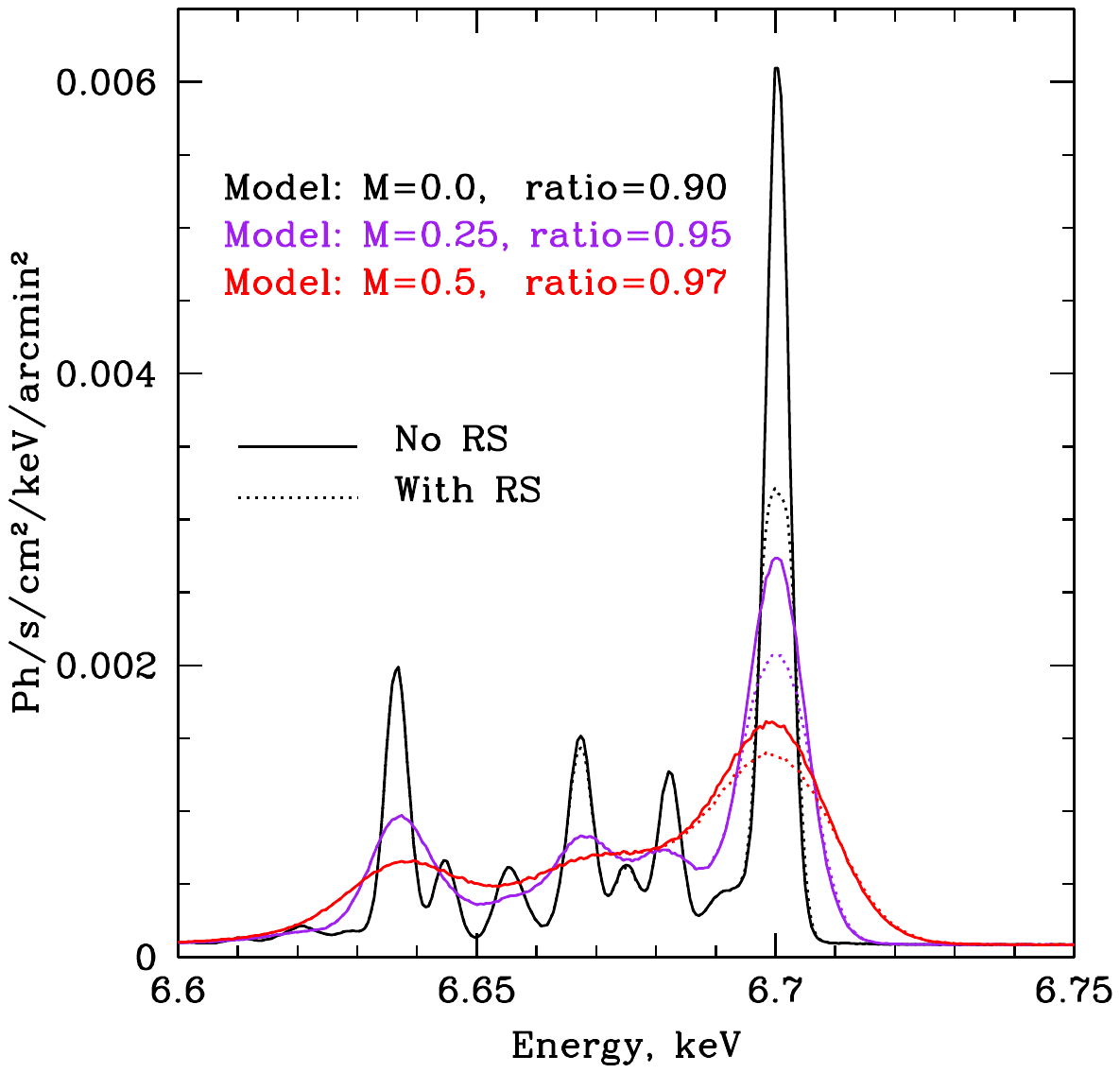}
\end{minipage}
\caption{
Predicted spectrum of the Perseus Cluster (from 0.5-2 arcmin annulus) with and without resonant scattering and different levels of gas turbulence.
The He-like Fe triplet line complex is shown for the Mach numbers 0, 0.25 and 0.5. Solid: no resonant scattering, dotted: with resonant scattering.
 The ratio of fluxes with and
without scattering integrated over 6.6-6.75 keV band is
shown in the left corner. As expected, the higher the velocity of the turbulent motions of gas, the broader lines and the weaker line distortions due to the resonant scattering.
\label{fig:mach}
}
\end{figure}

\section{Results}
\subsection{Spectral models of RS in Perseus}
\label{sec:results}

Fig. \ref{fig:rs_lines} and \ref{fig:rs_lines_6.7} show the results of our
simulations. We show the spectrum only for the 0.5-2 arcmin
annulus (see Fig. \ref{fig:perim}), where line suppression due to RS
is expected to be the strongest in the Perseus Cluster \citep[for more details
see][]{Chu04}. The central 0.5 arcmin region is excluded from the analysis
since there is a significant central AGN contribution. The top panel
in Fig. \ref{fig:rs_lines} shows model spectra with and without the RS. The corresponding optical depth as a function of energy is shown
in the bottom panel. Notice, that several lines in the 6.5-8.4 keV
band are affected by scattering. Seven lines have the optical depth
$\tau>0.1$, while the line of the He-like iron (K$_{\alpha}$ FeXXV with
the rest energy 6.7 keV) has the highest optical depth
$\tau\sim2$. Fig. \ref{fig:rs_lines_6.7} shows the same model spectra and
optical depths in closeup in the region at the He-like Fe line.  Table \ref{tab:strong_lines} lists the line energy,
transition information and the radial optical depth for the strongest lines obtained assuming ``free'' metal abundance (eq. \ref{eq:freeab})  and "peaked'' (eq. \ref{eq:flatab}) abundance models. As expected, the larger the optical depth, the stronger
 the line suppression due to the RS. For the K$_{\alpha}$ FeXXV
line at 6.7 keV, the suppression is $\sim 25$ per cent. Two other lines of K$_\beta$
FeXXV and K$_\alpha$ FeXXVI with the optical depths $\sim 0.4$ and $\sim
0.3$, respectively, have a line flux suppression of $\sim 5 $ per cent.

Varying the level of turbulence through the Mach number, we produce models that can be used to constrain the
amplitude of turbulent motions. Fig. \ref{fig:mach} shows the spectral
models in the 6.7 keV region obtained for the Mach numbers 0, 0.25 and
0.5. The higher the turbulence, the less prominent the RS and the
broader lines are. Integrating fluxes with and without scattering over
the line complex around 6.7 keV, we see that the suppression of the total
flux of the 6.7 keV complex decreases from 10 per cent at $M=0$ down to 3 per cent 
at $M=0.5$.

In order to simplify the comparison of our models with
observed data, we have tabulated model spectra in the XSPEC 
format\footnote{Models for the Perseus Cluster and brief description of how to use them can be downloaded at http://www.mpa-garching.mpg.de/$\sim$izhur/rs\_models\_perseus/}.

\begin{figure*}
\begin{minipage}{0.49\textwidth}
\includegraphics[trim=30 150 50 0,width=0.93\textwidth]{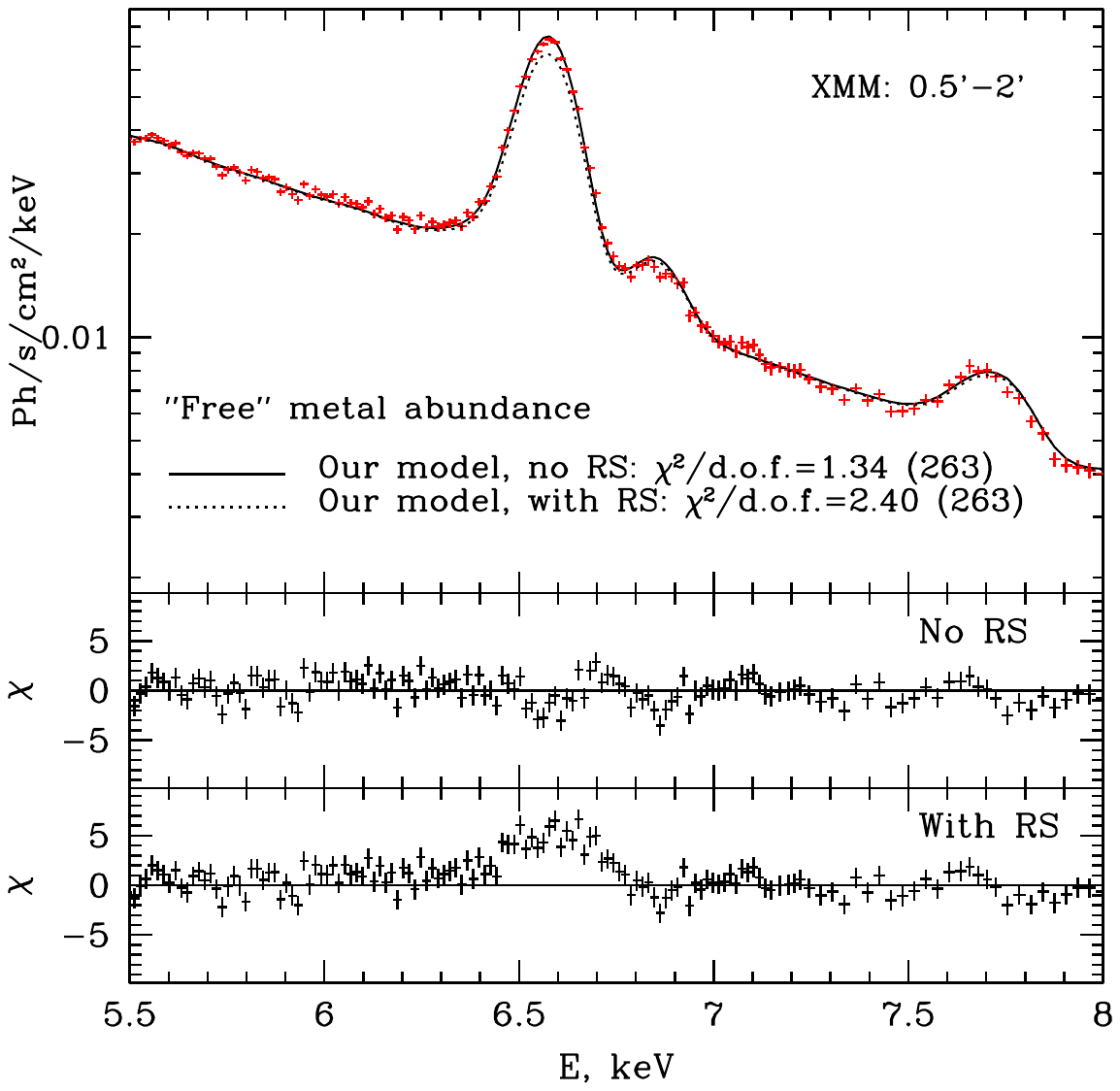}
\end{minipage}
\begin{minipage}{0.49\textwidth}
\includegraphics[trim=0 150 80 0,width=0.93\textwidth]{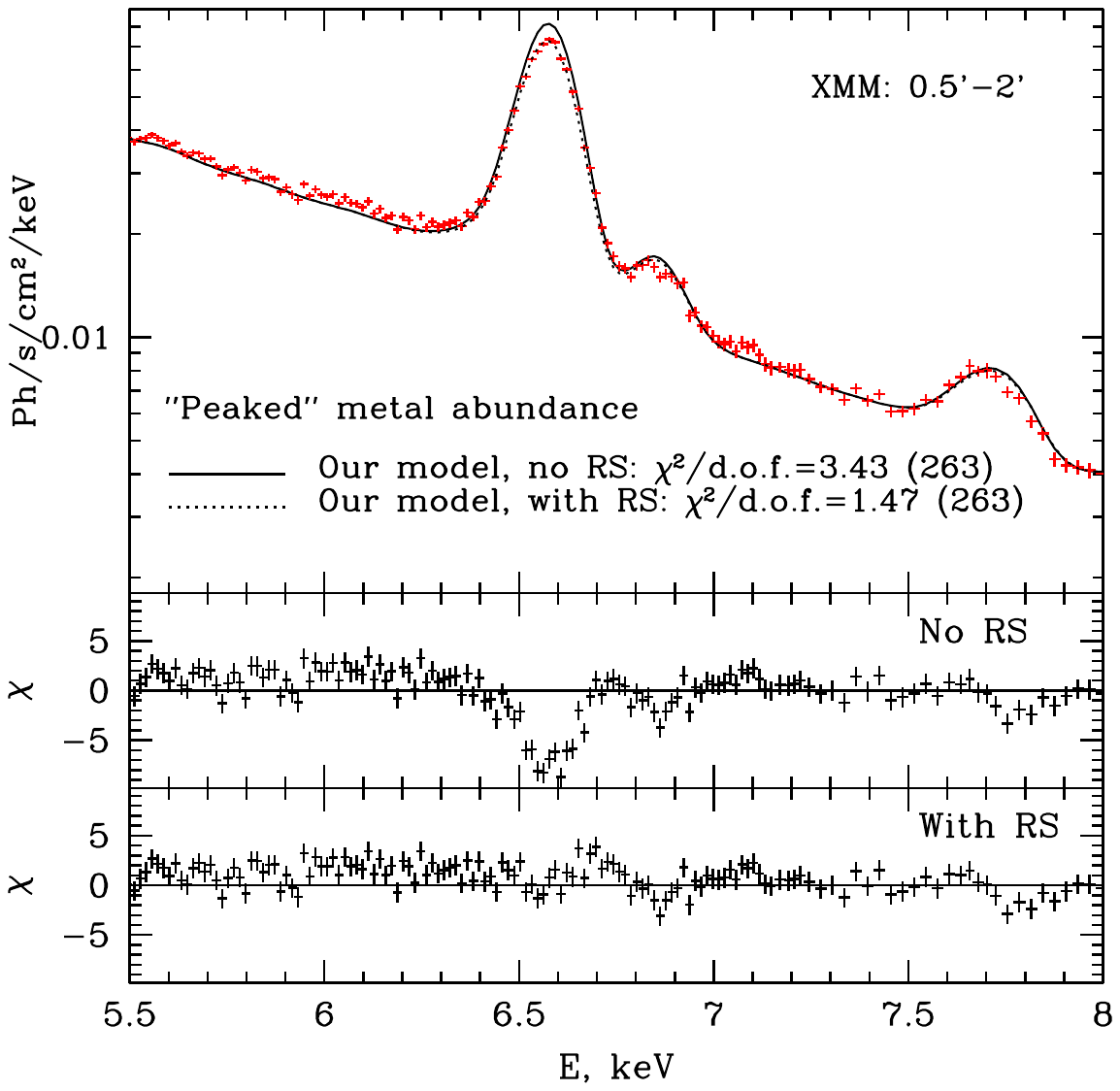}
\end{minipage}
\caption{
Comparison of our models with the {\it XMM-Newton} projected spectrum of the Perseus Cluster. Red points show the observed projected spectra from the central 0.5-2 arcmin annulus obtained from the {\it XMM-Newton} observations. Our spectral models are shown with black curves.  Solid/dotted curves correspond to the cases without/with resonant scattering. Deviations of our models from the observed data are shown in the lower panels. Here, we show models without turbulent motions of gas, the Mach number $M=0$. $\chi ^2$ per d.o.f. are labeled with a number of d.o.f. in brackets. {\bf Left:} {\it XMM} data and our model with a ``free'' abundance of heavy elements (eq. A3).  As can be seen from the figure, deviations of the model {\it with} resonant scattering are significantly stronger around 6.7 keV line complex. {\bf Right:} {\it XMM} data and results of our simulations with ``peaked'' metal abundance (eq. A4). Notice, that the energy resolution of current X-ray observatories does not allow us to separate spectral modifications due to different metal abundance and presence of the resonant scattering (see Fig. \ref{fig:mod_ab_mach} for more details).
\label{fig:rs_data_mod}
}
\end{figure*}

\begin{figure*}
\begin{minipage}{0.49\textwidth}
\includegraphics[trim=25 170 0 100,width=1.\textwidth]{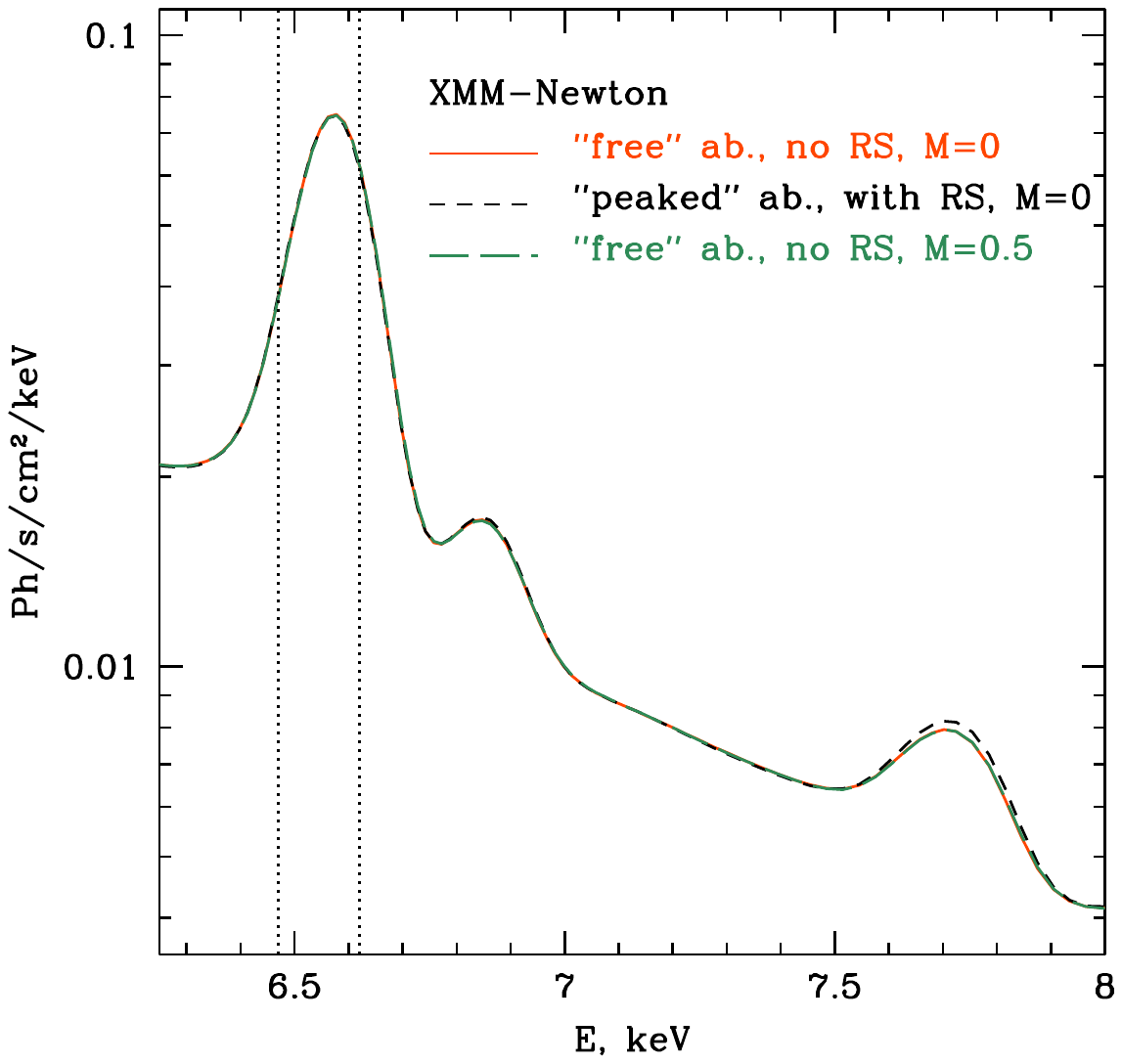}
\end{minipage}
\begin{minipage}{0.49\textwidth}
\includegraphics[trim=0 170 25 100,width=1.\textwidth]{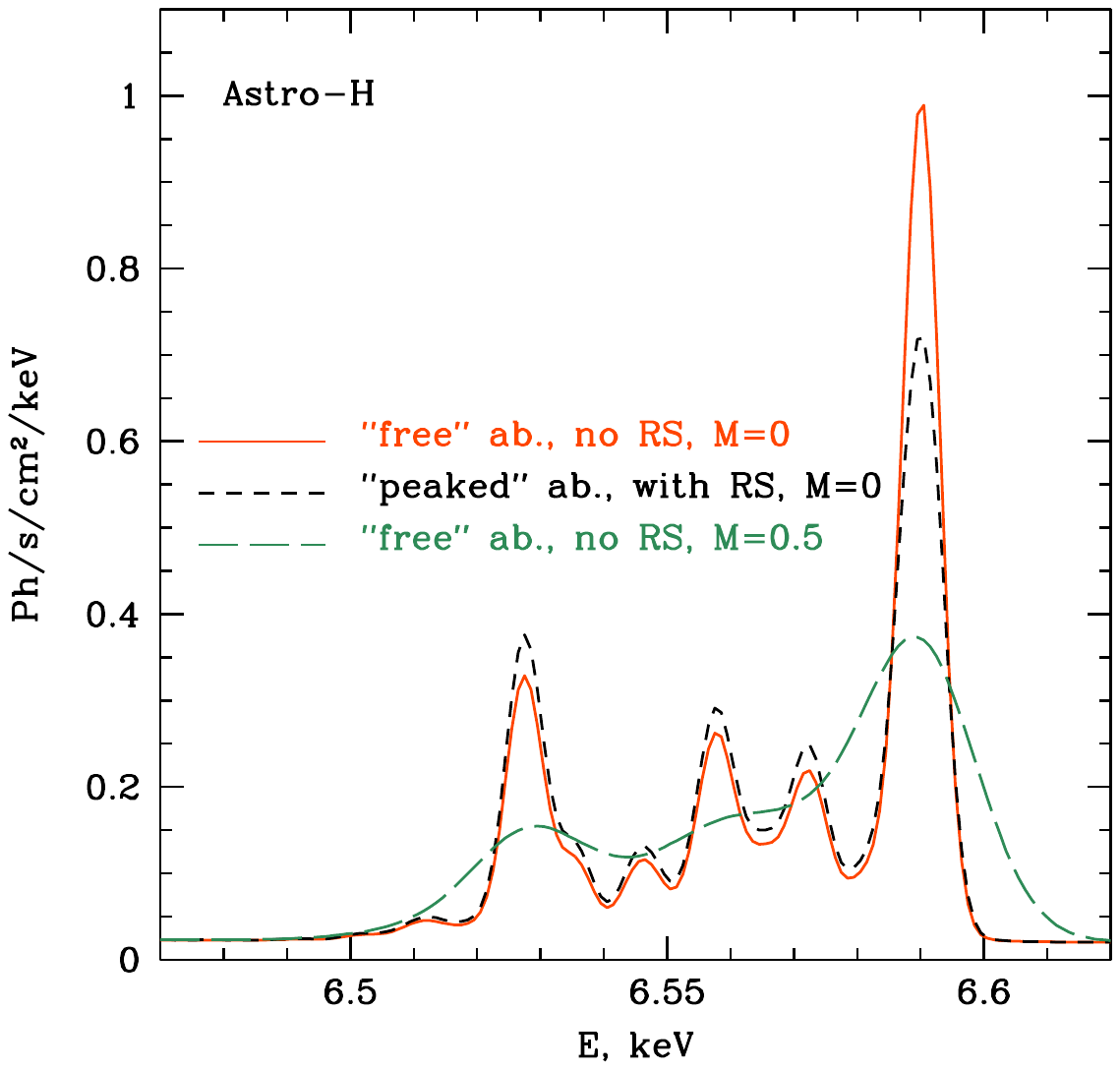}
\end{minipage}
\caption{Sensitivity of predicted spectra convolved with response matrices of the {\it XMM-Newton} (left) and {\it Astro-H} (right) observatories to the heavy elements abundance, the level of gas turbulence and presence of resonant scattering in the Perseus Cluster. The spectra for the 0.5-2 arcmin annulus are shown. Note, that the energy resolution of current X-ray instruments is not sufficient for resolving individual lines. Therefore, spectral distortions due to the resonant scattering, different metal abundance profiles and different level of gas turbulence are able to compensate each other in the final spectrum. In contrast, energy resolution of {\it Astro-H} is sufficient to detect and separate spectral distortions of different origin. Even a very narrow spectral region around 6.7 keV line (also indicated with a vertical dotted line in the left panel) allows us to do it.   
\label{fig:mod_ab_mach}
}
\end{figure*}

\subsection{Comparison with existing data}
\label{sec:compar}
 Below we compare our models with the 
existing {\it XMM-Newton} data on the Perseus Cluster. By default we compare data with models that have zero Mach number (no gas motions), unless other model is specified.

We use the projected spectrum from the 0.5-2 arcmin annulus, which is plotted with red points in Fig. \ref{fig:rs_data_mod}. The energy resolution of {\it XMM} is insufficient to resolve individual lines seen in the theoretical models (c.f. Fig. \ref{fig:rs_lines}). Instead, we see three bumps, which contain complexes
of lines: the $K_{\alpha}$ FeXXV complex around 6.7 keV (redshifted to $\sim$ 6.6 keV), the $K_{\alpha}$ FeXXVI complex at 6.97 keV (redshifted to $\sim$ 6.85 keV), and the $K_{\beta}$ FeXXV with $K_{\alpha}$ NiXXVII line complexes around 7.85 keV (redshifted to $\sim$ 7.7 keV). 

Given the uncertainty in the metal abundance profiles, we have
calculated two models assuming ``free" and ``peaked" abundance of
heavy elements (see Section \ref{sec:mod}). Models, convolved with
the {\it XMM-Newton} response are shown with the black curves on top of
the observed points in Fig. \ref{fig:rs_data_mod}. When doing the
comparison of the predicted and observed spectra, we do rudimentary
correction of the normalization (within 1-2 per cent) and the
redshift (less than $10^{-3}$) to ensure that the comparison is
dominated by the flux ratio of different lines rather than by the
difference in the observed and adopted values of the redshift and
density normalization.  The ``free" abundance model (left panel)
represents our best efforts to describe radial variations of the
observed metal abundance profile (derived from 1T fits of deprojected
spectra in the 5-9 keV band) under the assumption of optically thin
plasma. Notice, that in this case the model {\it without} the RS describes
data the best. The $\chi^2$ is reasonable, given the simplicity of the
model. It is only slightly higher than direct fitting of the projected
spectrum with the 1T APEC model. The fact that the 1T APEC model
marginally wins over our model is
not entirely surprising, since
deprojected profiles are designed to reproduce overall structure of
the cluster, rather than the projected spectrum in this particular annulus. The model with RS predicts
a line flux value well below the measured one (dotted curve in the left
panel of Fig. \ref{fig:rs_data_mod}). In contrast, the ``peaked" abundance
model is the result of several trials to find a radial metal abundance
profile, for which the model {\it with} the RS predicts a 6.7 keV line
flux broadly compatible with the data (the right panel). Indeed, we see that
in this case the model without the RS predicts significantly higher flux at
the 6.7 keV line complex.  The ``peaked" model is also driven by the
desire to have the abundance rising towards the center, rather
declining, like in the ``free" abundance model. To this end, the ``peaked"
model is one of the possible models and is not necessarily the very
best one in terms of $\chi^2$. Nevertheless, our ``peaked" model gives
$\chi^2$ just slightly higher than the case of ``free" metal
abundance.

 Comparison of our models with the {\it Chandra} data of the Perseus Cluster gives higher $\chi^2$ than for the {\it XMM-Newton} data. This is due to stronger deviations of the Ni-line complex from the model, as discussed in Section \ref{sec:mod}. In the case of ``free" abundance model, $\chi^2$/d.o.f. is 2.87 (d.o.f.=271) and 5.48 (271) for models without and with the RS, respectively. For the ``peaked" abundance, $\chi^2$/d.o.f. is 6.55 (271) and 2.90 (271) correspondingly.

Fig. \ref{fig:mod_ab_mach} compares the ``best-fitting" models discussed
above to model with ``free" metal abundance and turbulent motions of
the gas with Mach number M=0.5 (the left panel). We see that all three models
are barely distinguishable. The spectral resolution of the {\it
  XMM-Newton} (and {\it Chandra}) observatory is not sufficient to
robustly separate spectral distortions induced by the RS, different metal
abundance profiles and different levels of the gas turbulence. The situation
changes dramatically if we have {\it Astro-H} energy resolution $\sim$
5 eV. The right panel shows how differently spectra look if we vary
the abundance profile, presence/absence of the RS and the level of
turbulence. Notice, that with the {\it Astro-H} data we would be able to
distinguish different models and make conclusion on the RS effect in
Perseus even using a narrow energy band around the 6.7 keV line complex.

\subsection{Future prospects with {\it Astro-H}}

A new era of X-ray spectroscopy employing X-ray calorimetry will start
with the launch of Japanese mission {\it
Astro-H}\footnote{http://astro-h.isas.jaxa.jp/} \citep[see
e.g.][]{Tak10}. The Soft X-ray Spectrometer (SXS) system on-board of
{\it Astro-H} will 
have an energy resolution $\sim$ 5 eV between 0.3-10 keV, which is $\sim
30$ times better than the energy resolution around 6.7 keV line of the current X-ray CCDs. Such high spectral resolution is enough to resolve
individual lines in spectra of galaxy clusters. It will allows us to resolve the degeneracy between ``peaked abundance - RS" and ``free abundance - no RS" problem, mentioned in the previous Section. Also, looking at individual lines, the existing plasma models can be tested in great detail.

\begin{figure}
\begin{minipage}{0.48\textwidth}
\includegraphics[trim=20 155 20 100,width=1\textwidth]{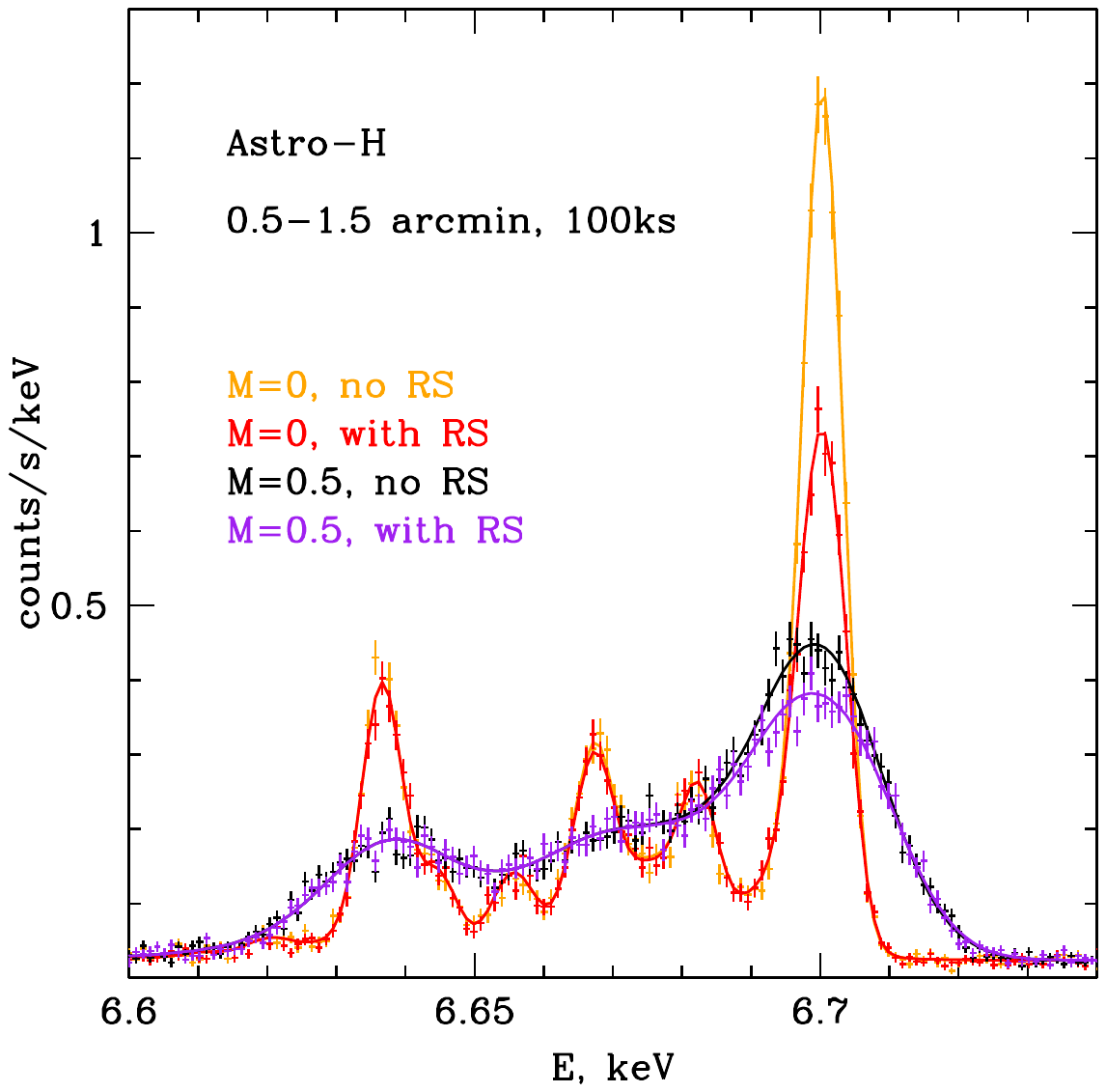}
\end{minipage}
\caption{
Simulated spectra from the core (0.5-1.5 arcmin annulus) of the Perseus
Cluster as {\it Astro-H} will obtain from 100 ks pointing observation. We used the
whole field of view (2.85 arcmin on a side) with excluded central chip, where the contribution of the central AGN is dominating the signal. Points show the ``observed'' spectra
(only statistical uncertainties are included), while solid curves show model spectra without and with resonant scattering respectively. Notice, that 100 ks observation is enough to detect the resonant scattering signal even if motions of gas are present.
\label{fig:astroh}
}
\end{figure}

The effect of the RS in the core of the Perseus Cluster will be readily seen already
with 100 ks observations. Fig. \ref{fig:astroh} shows simulated
spectra from the 0.5-1.5 arcmin annulus in Perseus as {\it Astro-H} will
measure with a 100 ks pointing observation. We used the
whole field of view, 2.85 arcmin on a side, with excluded central chip, where the contribution of the central AGN is dominating the signal. The RMF and ARF files were taken from http://astro-h.isas.jaxa.jp/. We used the SXS energy response assuming constant energy resolution of 5 eV, while the ARF for all pixels was taken for point sources with open filter. Simulations neglect the redistribution of photons due to the mirror point-spread-function. Points show simulated spectra with and without RS for Mach
numbers 0 and 0.5, while solid curves show our models, convolved
with the {\it Astro-H} response. Notice, even if the level of
turbulence is high (M=0.5), {\it Astro-H} will be easily capable of measuring 
the suppression of lines. Here we assumed only statistical
uncertainties.  The next step will be to perform more realistic
simulations of the spectra in the Perseus Cluster, accounting for mirror
blurring and the finite pixel size of the detector (Konami et al. 2013, in prep.).

Note that the above model was produced assuming small-scale gas
motions. In this context small-scale means scales much smaller than
the characteristic size of the Perseus Cluster core. This implies that
a Gaussian is a good description of the line profile. When large scale
motions are present the line shape may have a complicated shape
dominated by stochastic velocity fields \citep{Zhu12,Sha12}. Useful approaches for studying such cases may include the use of optically thin lines to infer template line shapes
 or the approaches developed by
\citet{Sha12}. Note that the amplitude of the RS effect
depends on the dominant anisotropy in the velocity field
\citep{Zhu11}. It should be possible therefore to test
if the line broadening/shape and the line ratios are self-consistent.

\section{Conclusions}
We have produced spectral models for the Perseus Cluster which take
into account modifications of line fluxes and shapes due to the RS
effect, line broadening by small-scale motions and radial variations of
density, temperature and abundance of heavy elements. Similar
calculations can be easily extended for any spherically-symmetric
model of other clusters. The main advantage of the model is that it
self-consistently accounts for projection effects, line broadening and scattering of photons produced not only for a single line,
but for the whole spectrum at once. This is particularly important when line
broadening causes overlap of individual lines.  Our models can be
directly compared to data in a certain energy band, accounting for
the suppression of all optically thick lines in this band. This provides more robust constraints on the RS and the amplitude of gas motions in
clusters. The model spectra are stored in the XSPEC format and can be
used straightforwardly in a usual way.

We have compared our model to the existing data of the Perseus Cluster
from the {\it Chandra} and {\it XMM-Newton} observatories. We do not see
strong evidence for the RS in the central 0.5-2 arcmin region. However,
we showed that a firm conclusion on the presence or absence of the RS
cannot be easily made with existing data since the limited energy
resolution of the spectra complicates model comparisons. Spectral
models convolved with the CCD resolution are largely degenerate if we
vary the metal abundance profile and/or the Mach number. With the current data one has to
rely on the comparison of line fluxes from different ions or elements.

The high energy resolution of the future X-ray observatory {\it Astro-H} should
enable us to resolve most of the problems we are facing with the current
data. The abundance of heavy elements will be better constrained since
we will see individual lines of the same ions in spectra. Calculations
of the line emissivities using APEC models of optically thin plasma  (or any other plasma emission code) will be tested on resolved optically
thin lines in spectra from the {\it Astro-H} observations. Modifications
in spectra due to the presence of turbulence will be easy to
detect. Therefore, analysis of the RS effect can be done
self-consistently with the help of models presented in this work. We
have shown that 100 ks observations will be enough to make a
significant step forward in the RS measurements.

\section{Acknowledgements} 
 This work is supported in part by the U.S. Department of Energy under
  contract number DE-AC02-76SF00515. The research made use of grant NSh-5603.2012.2, programs P-19 and OFN-17 of the Russian Academy of Sciences, and RFBR grant 13-02-01365.

\appendix
\section{Spherically-symmetric model of the Perseus Cluster}
\label{sec:appen}
Below we summarize our model of the Perseus Cluster used as an input to our simulations of spectra with RS (see Fig. \ref{fig:deproj}). The number electron density [cm$^{-3}$] radial profile is
\beq
&\nonumber n_e=\left[\disp\frac{4.6\cdot 10^{-2}}{\left(1+\disp\left(\frac{r}{55}\right)^2\right)^{1.8}}+\disp\frac{4.8\cdot 10^{-3}}{\left(1+\left(\disp\frac{r}{200}\right)^2\right)^{0.87}}\right]\times&\\&\times\disp\left[1-0.06\exp \left(-\left(\disp\frac{r-30}{9}\right)^2\right)\right]\times&\\ &\nonumber \times \disp\left[1+0.04\exp \left(-\left(\disp\frac{r-15}{8}\right)^2\right)\right]&,
\eeq
where $r$ is in kpc. The profile of temperature [keV] is
\be
T_e=7.5\disp\frac{1+\left(\disp\frac{r}{58}\right)^{3.5}}{2.45+\left(\disp\frac{r}{58}\right)^{3.6}}\times \disp\frac{1.55+\left(\disp\frac{r}{20}\right)^{2.04}}{1+\left(\disp\frac{r}{20}\right)^{2}}.
\ee
"Free'' metal abundance profile (relative to $Z_{\odot}$) is
\be
Z=0.33\disp\frac{2+1.1\left(\disp\frac{r}{90}\right)^{2.7}}{1.1+\left(\disp\frac{r}{90}\right)^{2.7}}\times \disp\frac{0.73+\left(\disp\frac{r}{30}\right)^{3}}{1+\left(\disp\frac{r}{30}\right)^{3}},
\label{eq:freeab}
\ee
while "peaked'' abundance is assumed the following
\be
Z=0.355\disp\frac{2+1.1\left(\disp\frac{r}{70}\right)^{2.7}}{1.1+\left(\disp\frac{r}{70}\right)^{2.7}}\times \disp\frac{1+\left(\disp\frac{r}{1.5}\right)}{0.8+\left(\disp\frac{r}{1.5}\right)}.
\label{eq:flatab}
\ee
\label{lastpage}
\end{document}